\pdfoutput=1  
\documentclass{bmvc2k}

\title{Domain-Aware Lightweight Spectral-Grouped Convolutions for Hyperspectral Fish Freshness Classification}

\addauthor{Kazi Nabiul Alam}{K.Alam3742@student.leedsbeckett.ac.uk}{1}
\addauthor{Pooneh Bagheri Zadeh}{P.Bagheri-Zadeh@leedsbeckett.ac.uk}{1}
\addauthor{Akbar Sheikh-Akbari}{A.Sheikh-Akbari@leedsbeckett.ac.uk}{1}

\addinstitution{
School of Built Environment, Engineering, and Computing, \\ Leeds Beckett University.\\  Leeds LS6 3QS, United Kingdom. 
}

\runninghead{Alam et. al}{SGNet: Domain-Aware HSI Fish Freshness}

\def\etal{\emph{et al}\bmvaOneDot}

\usepackage{algorithm}
\usepackage{algpseudocode}
\usepackage{amsmath}
\usepackage{amssymb}   

\setcitestyle{numbers,square,comma}

\begin{document}
\maketitle

\begin{abstract}
Hyperspectral imaging (HSI) offers nondestructive assessment of fish freshness by detecting biochemical alterations across spectral bands. However, conventional deep learning approaches do not fully address the particular characteristics of HSI data, such as spectral dominance over spatial textures, ordinal label structure, and a small number of training samples. We propose SGNet (Spectral-Grouped Network), a lightweight architecture that separates spectral and spatial feature extraction using grouped convolutions and a depthwise spatial pathway. A dual attention mechanism that couples channel-wise squeeze-and-excitation with spatial gating adaptively highlights informative features. SGNet achieves 97.8\% classification accuracy and 0.64 days mean absolute error (MAE) with just 4.75M parameters when tested on our newly developed 16-day refrigerator-stored salmon fillet dataset. Ablation studies validate the contribution of each component, while comparisons demonstrate a five- to eighteen-fold parameter reduction relative to ResNet-50 and Vision Transformers. Our findings indicate that domain-aware design supports precise, real-time freshness prediction for industrial implementation.
\end{abstract}

\section{Introduction}
\label{sec:intro}

Fish freshness is a significant determinant of nutritional content, consumer safety, and market value in the global seafood industry~\cite{c1, c2}. Conventional assessment techniques, such as sensory evaluation, microbiological testing, and chemical analysis, are variously subjective, destructive, or time-intensive, which makes them inappropriate for real-time industrial use~\cite{c3, c4}. The growing body of research into sensing technologies reflects a clear demand for automated, non-destructive, and rapid quality inspection.

Hyperspectral imaging (HSI) has emerged as an effective tool for evaluating food quality by integrating spatial imaging with extensive spectral data acquired across numerous contiguous wavelength bands~\cite{c5, c6}. An HSI system operates differently from standard RGB imaging because it is able to identify chemical breakdown products through the analysis of specific spectral bands~\cite{c7, c8}. These traits make HSI especially useful for estimating fish freshness, since the quality of fish degrades gradually and continuously over time, leaving a measurable spectral signature at each stage of spoilage.

Although classical machine learning methods such as partial least squares discriminant analysis (PLS-DA) and support vector machines (SVM) have demonstrated potential for HSI-based freshness evaluation~\cite{c9, c10}, they depend heavily on manually designed features and prior wavelength selection~\cite{c11}, which limits their capacity to capture detailed spectral-spatial interactions. Deep learning techniques have achieved superior results in hyperspectral image classification according to recent research~\cite{c12}, yet most established models were designed for remote sensing tasks. Such models do not adequately accommodate the characteristics of food quality data, which combine large spectral dimensionality with restricted sample availability and an inherently ordinal label structure. A further difficulty, frequently overlooked, is the substantial redundancy between neighbouring spectral bands; indiscriminate mixing of all bands therefore risks entangling correlated and unrelated wavelength responses, which amplifies noise and encourages overfitting under limited data regimes.

This study introduces SGNet (Spectral-Grouped Convolutional Neural Network), a framework designed around the structure of hyperspectral food data rather than adapted from a remote sensing or natural image backbone. Rather than proposing new primitives, our contribution is a principled composition of efficient operators that jointly respects spectral dominance, ordinal label structure, and severe sample scarcity, together with the analysis and dataset needed to validate it in this regime. Our contributions are as follows:

\begin{itemize}
    \item An explicit spectral-spatial factorisation in which grouped pointwise convolution confines cross-band interaction to local channel subsets while a parallel depthwise pathway captures spatial structure, preventing premature entanglement of unrelated wavelength responses.
    \item A lightweight dual attention mechanism coupling channel-wise excitation with spatial gating at below 5\% of the block parameters, giving selective emphasis without the data appetite of self-attention.
    \item A hierarchical encoder, formalised in Algorithm~\ref{alg:sgnet}, that attains 97.8\% accuracy with only 4.75M parameters and is benchmarked against widely used convolutional and transformer architectures retrained under identical conditions.
    \item A curated 16-day refrigerated salmon dataset with strict pack-level separation, addressing the scarcity of leakage-controlled hyperspectral freshness data, which we release to support reproducibility.
\end{itemize}

\section{Related Work}
\label{sec:related}

\subsection{Hyperspectral Imaging for Fish Freshness}

HSI has been applied extensively to fish freshness assessment in recent years. Cheng~\etal~\cite{c1} employed HSI with least squares SVM to predict total volatile basic nitrogen (TVB-N) content in grass carp, while Yu~\etal~\cite{c2} combined HSI with data fusion for rapid evaluation of tilapia fillet freshness. Falahatnejad~\etal~\cite{c3} surveyed fish quality assessment using HSI and computer vision, distinguishing the conventional pipelines that dominate the field. Hardy~\etal~\cite{c13} applied K-nearest neighbours to hyperspectral data for salmon storage-day prediction, achieving 77\% accuracy. Kashani Zadeh~\etal~\cite{c14} developed a multimode spectroscopic system that combined VIS-NIR, SWIR, and fluorescence data for salmon freshness classification, and Khoshnoudi-Nia and Moosavi-Nasab~\cite{c4} predicted multiple freshness indicators in fish fillets from a single multispectral imaging system. These studies establish that HSI data carry rich freshness information, yet they predominantly rely on conventional machine learning pipelines that require human intervention to engineer features and select informative wavelengths.

\subsection{Deep Learning for Hyperspectral Classification}

Deep learning has transformed hyperspectral image analysis through its capacity for hierarchical feature extraction. Early CNN-based methods~\cite{c15, c16} used 3D convolutional layers to fuse spectral and spatial information jointly, but their high computational requirements limited their practicality. Roy~\etal~\cite{c17} subsequently proposed HybridSN, a hybrid 3D--2D convolutional design that reduces the cost of pure 3D models while retaining strong spectral-spatial discrimination. In parallel, attention mechanisms have proven highly effective: Hu~\etal~\cite{c18} introduced squeeze-and-excitation (SE) blocks for channel recalibration, and Woo~\etal~\cite{c19} proposed CBAM, which unifies channel and spatial attention. For HSI specifically, spectral and recurrent attention networks~\cite{c20, c12} have achieved state-of-the-art results on remote sensing benchmarks. A recurring limitation is that these systems are calibrated for large-scale data collections and incur considerable computational cost, which is poorly matched to the small-sample, high-dimensional regime that characterises food quality inspection.

\subsection{Efficient Architectures and Ordinal Modelling}

Advances in efficient network design provide the primitives on which SGNet builds. Howard~\etal~\cite{c21} introduced depthwise separable convolutions in MobileNet, reducing computation by roughly eight to nine times relative to standard convolution. Tan and Le~\cite{c22} formalised compound scaling in EfficientNet to balance depth, width, and resolution. Liu~\etal~\cite{c23} revisited the convolutional design space with ConvNeXt and showed that pure ConvNets can rival Vision Transformers~\cite{c24}, while hierarchical transformers such as Swin~\cite{c25} brought locality and multi-scale structure to attention-based models. Grouped convolutions, popularised by ResNeXt~\cite{c26}, occupy a middle ground between standard and depthwise convolution by partitioning channels into groups. A complementary line of work addresses the ordinal nature of many prediction tasks: Cao~\etal~\cite{c27} proposed the rank-consistent CORAL framework for ordinal regression, and Lin~\etal~\cite{c28} introduced focal loss to address class imbalance, both of which are directly relevant to freshness estimation where labels evolve continuously over time. SGNet draws on these efficient primitives but arranges them according to the specific structure of hyperspectral food data rather than transplanting a remote sensing or natural image backbone unchanged.

A concise comparison of representative approaches is given in Table~\ref{tab:related}, which contrasts the input modality, the feature extraction strategy, the treatment of label ordinality, and the relative parameter scale. The table makes clear that prior HSI freshness work tends either to rely on hand-crafted features or to inherit heavyweight backbones, leaving a gap for a lightweight, domain-aware design that respects spectral dominance and ordinal structure simultaneously.

\begin{table*}[t]
\centering
\caption{Comparison of representative approaches to HSI-based food quality assessment and related deep architectures. ``Hand-crafted'' denotes manual feature or wavelength selection; ``Eval.'' denotes ordinality handled at evaluation time; ``--'' denotes not applicable.}
\label{tab:related}
\setlength{\tabcolsep}{0.9pt}
\begin{tabular}{lcccc}
\hline
Method & Input & Feature Strategy & Ordinal-Aware & Scale \\
\hline
Classical SVM~\cite{c1,c10} & HSI / spectra & Hand-crafted & No & -- \\
KNN salmon~\cite{c13} & HSI & Hand-crafted & No & -- \\
3D-CNN~\cite{c15,c16} & HSI cube & Learned (3D) & No & Heavy \\
HybridSN~\cite{c17} & HSI cube & Learned (3D--2D) & No & Medium \\
SE / CBAM~\cite{c18,c19} & Image & Learned + attention & No & Medium \\
SpectralFormer~\cite{c12} & HSI & Transformer & No & Heavy \\
ConvNeXt~\cite{c23} & Image & Learned (ConvNet) & No & Heavy \\
\textbf{SGNet (Ours)} & HSI cube & Grouped + DW + attention & Eval. & \textbf{Light} \\
\hline
\end{tabular}
\end{table*}

\section{Methodology}
\label{sec:methodology}

\subsection{Problem Setting and Notation}

Let $\mathcal{D} = \{(X_i, y_i)\}_{i=1}^{N}$ denote a hyperspectral fish freshness dataset, where $X_i \in \mathbb{R}^{C \times H \times W}$ represents a hyperspectral image cube with $C$ spectral bands, and $y_i \in \{0,1,\dots,K-1\}$ denotes the freshness day label. Unlike standard image classification, the labels are ordinal, reflecting the continuous temporal evolution of freshness rather than a set of mutually independent categories.

The objective is to learn a mapping
\begin{equation}
f_\theta: \mathbb{R}^{C \times H \times W} \rightarrow \mathbb{R}^{K},
\end{equation}
in which prediction errors are penalised not only in a categorical sense but also with respect to the ordinal distance between predicted and true freshness days~\cite{c27}.

\begin{figure*}[t]
  \centering
  \includegraphics[width=\textwidth]{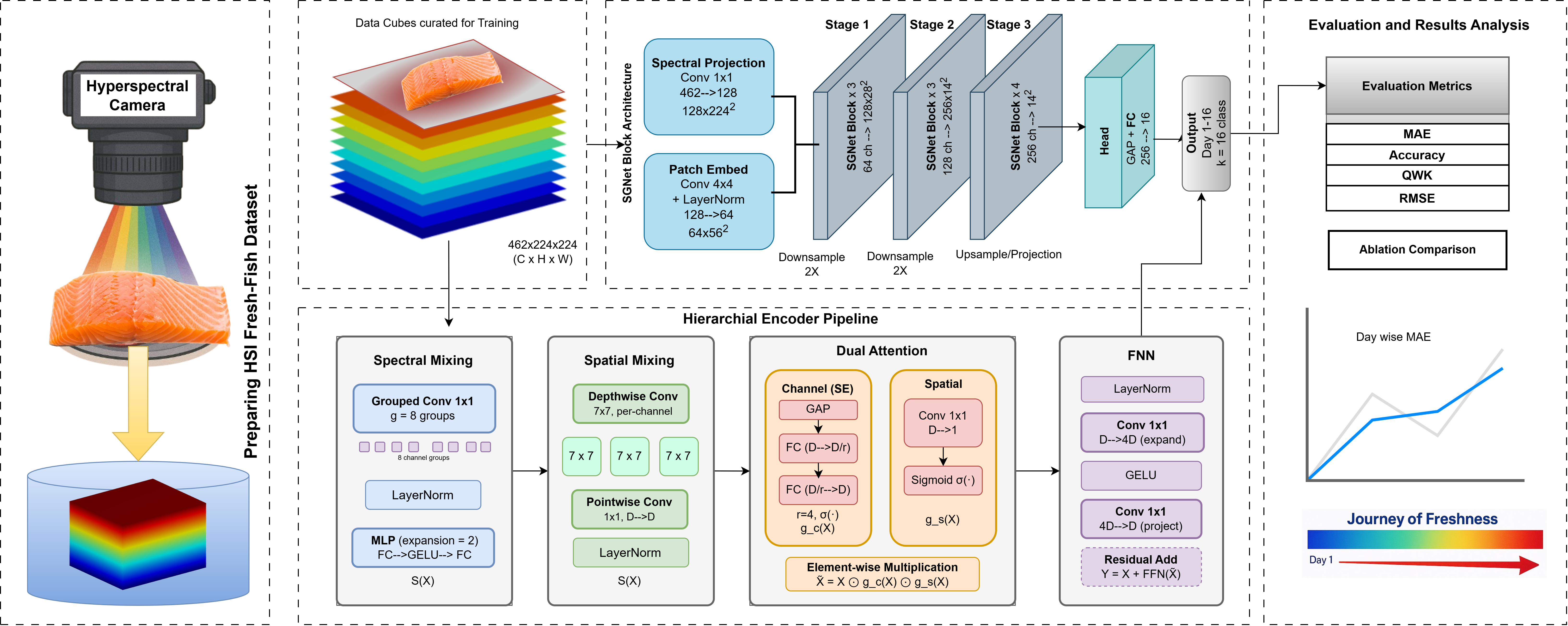}
  \caption{Proposed SGNet framework for hyperspectral fish freshness classification.}
  \label{fig:1}
\end{figure*}

\subsection{Design Rationale}

The hyperspectral fish freshness classification task presents three principal obstacles. First, spectral dominance means that freshness information is carried primarily by wavelength-dependent reflectance patterns rather than by spatial texture. Second, temporal smoothness requires that representations of neighbouring days remain close in feature space, so that an error of one day is treated as far less severe than an error of several days. Third, the available data are limited in number and are subject to additional sources of variation and noise. Generic CNNs that combine spectral and spatial information without regard for these constraints perform poorly, because they entangle correlated bands prematurely and overfit to spurious patterns. The proposed design therefore contains two independent processing paths that use grouped convolutions together with lightweight attention to suppress unimportant variation while performing spectral and spatial interaction in a controlled manner.

Table~\ref{tab:architecture} and Figure~\ref{fig:1} summarise the SGNet architecture.

\begin{table}[t]
\centering
\caption{SGNet Architecture Overview}
\label{tab:architecture}
\begin{tabular}{llcc}
\hline
Stage & Layers & Output Size & Params \\
\hline
Stem & Spec. Proj. + Patch Emb. & $64 \times 56^2$ & 74K \\
Stage 1 & SGNet $\times 3$ + Down & $128 \times 28^2$ & 429K \\
Stage 2 & SGNet $\times 3$ + Down & $256 \times 14^2$ & 1.31M \\
Stage 3 & SGNet $\times 4$ & $256 \times 14^2$ & 2.89M \\
Head & GAP + Linear & $K$ & 4.1K \\
\hline
Total &  &  & \textbf{4.75M} \\
\hline
\end{tabular}
\end{table}

\subsection{Spectral Projection and Patch Embedding}

We first perform a linear spectral projection with a point-wise convolution on the hyperspectral cube $X \in \mathbb{R}^{C \times H \times W}$:
\begin{equation}
X^{(0)} = \mathrm{LN}\left(W_0 * X\right), \quad W_0 \in \mathbb{R}^{C_0 \times C \times 1 \times 1},
\end{equation}
where $C_0=64$ and $\mathrm{LN}(\cdot)$ denotes channel-wise layer normalisation. This operation performs controlled spectral mixing, compressing the $C$-dimensional spectral signature into a learned $C_0$-dimensional embedding while preserving full spatial resolution.

To obtain spatially efficient representations, a strided convolution is subsequently applied:
\begin{equation}
X^{(1)} = \mathrm{LN}\left(W_1 * X^{(0)}\right), \quad W_1 \in \mathbb{R}^{64 \times 64 \times 4 \times 4},
\end{equation}
yielding feature maps of size $64 \times \frac{H}{4} \times \frac{W}{4}$. Unlike the pooling-based patchification common in vision transformers~\cite{c24}, this learnable downsampling preserves spectral fidelity while enabling early spatial abstraction.

\subsection{Spectral--Grouped CNN Block}

The core computational unit is the Spectral--Grouped CNN (SGNet) block, which processes spectral and spatial information through dedicated pathways. Let $X \in \mathbb{R}^{D \times h \times w}$ denote the input feature map. Each block comprises four sequential operations.

\subsubsection{Spectral Mixing}
Cross-channel interaction is performed via grouped pointwise convolution:
\begin{equation}
S(X) = \phi\left(\mathrm{LN}\left(W_s^{(g)} * X\right)\right),
\end{equation}
where $W_s^{(g)}$ denotes a $1\times1$ convolution with $g=8$ groups, $\phi(\cdot)$ is GELU activation, and the output is further processed by a two-layer MLP with expansion ratio 2. Grouped convolution constrains cross-band interaction to local channel subsets, preventing premature entanglement of unrelated spectral features and reducing sensitivity to noise.

\subsubsection{Spatial Mixing}
Local spatial context is aggregated independently per channel using depthwise separable convolution:
\begin{equation}
P(X) = W_p * \left(W_d * X\right),
\end{equation}
where $W_d \in \mathbb{R}^{D \times 1 \times k \times k}$ is a depthwise convolution with kernel size $k=7$, and $W_p \in \mathbb{R}^{D \times D \times 1 \times 1}$ is a pointwise projection. This factorisation captures local tissue structure and spatial patterns while preserving spectral independence.

\subsubsection{Dual Attention Gating}
To adaptively emphasise informative spectral responses and spatial regions, we introduce a lightweight dual attention mechanism inspired by squeeze-and-excitation~\cite{c18} and convolutional attention~\cite{c19}. Channel-wise attention is computed via squeeze-and-excitation:
\begin{equation}
g_c(X) = \sigma\left(W_{c2} \cdot \delta\left(W_{c1} \cdot \mathrm{GAP}(X)\right)\right),
\end{equation}
where $\mathrm{GAP}(\cdot)$ denotes global average pooling, $\delta(\cdot)$ is ReLU activation, and the bottleneck reduction ratio is $r=4$. Spatial attention is computed as:
\begin{equation}
g_s(X) = \sigma\left(W_a * X\right),
\end{equation}
where $W_a \in \mathbb{R}^{1 \times D \times 1 \times 1}$ produces a spatial attention map. The gated output combines both attention mechanisms:
\begin{equation}
\tilde{X} = X \odot g_c(X) \odot g_s(X),
\end{equation}
where $\odot$ denotes element-wise multiplication with appropriate broadcasting. Unlike self-attention, this mechanism introduces minimal computational overhead (less than 5\% of block parameters) and remains stable under limited data regimes.

\subsubsection{Feed-Forward Network}
The final output is obtained via a residual feed-forward transformation:
\begin{equation}
Y = X + \mathrm{FFN}\left(\mathrm{LN}(\tilde{X})\right),
\end{equation}
where $\mathrm{FFN}(\cdot)$ consists of two pointwise convolutions with expansion ratio 4 and GELU activation:
\begin{equation}
\mathrm{FFN}(Z) = W_2 * \phi(W_1 * Z), \quad W_1 \in \mathbb{R}^{4D \times D}, W_2 \in \mathbb{R}^{D \times 4D}.
\end{equation}

\subsection{Hierarchical Encoder}

The complete network stacks SGNet blocks across three stages with progressively increasing channel dimensions $\{64, 128, 256\}$ and correspondingly decreasing spatial resolutions $\{56^2, 28^2, 14^2\}$. Block depths are set to $\{3, 3, 4\}$ for stages 1--3 respectively. Downsampling between stages is performed via layer-normalised strided convolution:
\begin{equation}
X_{l+1} = \mathrm{Conv}_{2\times2,s=2}\left(\mathrm{LN}(X_l)\right).
\end{equation}
This hierarchical design enables the model to capture freshness-related cues across multiple spatial scales, ranging from fine-grained local tissue variation to coarse global structural change indicative of degradation. The progressive structure mirrors the multi-scale philosophy of hierarchical transformers~\cite{c25} while retaining the efficiency of convolutional primitives.

\subsection{Classification Head}

Global average pooling aggregates the final-stage features into a location-invariant representation:
\begin{equation}
z = \mathrm{GAP}(X_L) \in \mathbb{R}^{256}.
\end{equation}
A linear classifier produces logits for the $K$ ordinal freshness classes:
\begin{equation}
\hat{y} = W_o z + b_o, \quad W_o \in \mathbb{R}^{K \times 256}.
\end{equation}
The model is trained with standard cross-entropy loss, which we adopt as a deliberately simple and strong baseline; this isolates the effect of the architecture from that of the loss. We evaluate using ordinal-aware metrics including mean absolute error (MAE), accuracy (Acc), and quadratic weighted kappa (QWK), which better reflect the continuous nature of freshness evolution and penalise predictions in proportion to their distance from the ground truth. That the architecture alone yields strong ordinal behaviour under a non-ordinal loss is itself informative; we discuss the integration of an explicit rank-consistent objective~\cite{c27} as a direction for future work in Section~\ref{sec:discussion}.

\subsection{The SGNet Forward Pass}

For clarity, Algorithm~\ref{alg:sgnet} collects the operations defined above into a single end-to-end procedure. The listing makes explicit the order in which spectral and spatial information are processed and shows that the two pathways remain decoupled until the dual attention stage recombines them. We regard this controlled separation, rather than the early joint mixing favoured by generic spectral-spatial networks, as the principal source of the model's sample efficiency.

\begin{algorithm}[t]
\caption{SGNet forward pass for a hyperspectral cube.}
\label{alg:sgnet}
\begin{algorithmic}[1]
\Require Cube $X \in \mathbb{R}^{C \times H \times W}$; stage depths $\{n_1,n_2,n_3\}=\{3,3,4\}$; widths $\{64,128,256\}$
\Ensure Ordinal class logits $\hat{y} \in \mathbb{R}^{K}$
\State $X^{(0)} \gets \mathrm{LN}(W_0 * X)$ \Comment{spectral projection to $C_0{=}64$}
\State $X \gets \mathrm{LN}(W_1 * X^{(0)})$ \Comment{strided patch embedding, stride $4$}
\For{stage $s = 1$ to $3$}
    \For{block $= 1$ to $n_s$}
        \State $S \gets \phi(\mathrm{LN}(W_s^{(g)} * X))$ \Comment{grouped spectral mixing, $g{=}8$}
        \State $S \gets \mathrm{MLP}_{2}(S)$ \Comment{expansion ratio $2$}
        \State $P \gets W_p * (W_d * S)$ \Comment{depthwise spatial mixing, $k{=}7$}
        \State $g_c \gets \sigma(W_{c2}\,\delta(W_{c1}\,\mathrm{GAP}(P)))$ \Comment{channel attention}
        \State $g_s \gets \sigma(W_a * P)$ \Comment{spatial attention}
        \State $\tilde{X} \gets P \odot g_c \odot g_s$ \Comment{dual gating}
        \State $X \gets X + \mathrm{FFN}(\mathrm{LN}(\tilde{X}))$ \Comment{residual FFN, ratio $4$}
    \EndFor
    \If{$s < 3$}
        \State $X \gets \mathrm{Conv}_{2\times2,\,s=2}(\mathrm{LN}(X))$ \Comment{downsample}
    \EndIf
\EndFor
\State $z \gets \mathrm{GAP}(X)$ \Comment{location-invariant embedding}
\State $\hat{y} \gets W_o\,z + b_o$ \Comment{ordinal classification head}
\State \Return $\hat{y}$
\end{algorithmic}
\end{algorithm}

\section{Experiments}
\label{sec:experiments}

\subsection{Experimental Setup}

\textbf{Dataset and Implementation.} Our curated salmon hyperspectral dataset \cite{c29} consists of $N = 800$ HSI cubes acquired over a 16-day refrigerated storage period ($K = 16$ classes), where day 6 is the labelled expiry date. Each cube comprises $C = 462$ spectral bands spanning the visible and near-infrared range. The dataset is split into training (560), validation (112), and test (128) sets with balanced class distributions. To avoid data leakage, distinct fish packs were assigned to each split (35 packs for training, 7 for validation, and 8 for testing), so that no cube from a given pack appears in more than one split.

\begin{figure*}[htbp]
  \centering
  \includegraphics[width=\textwidth]{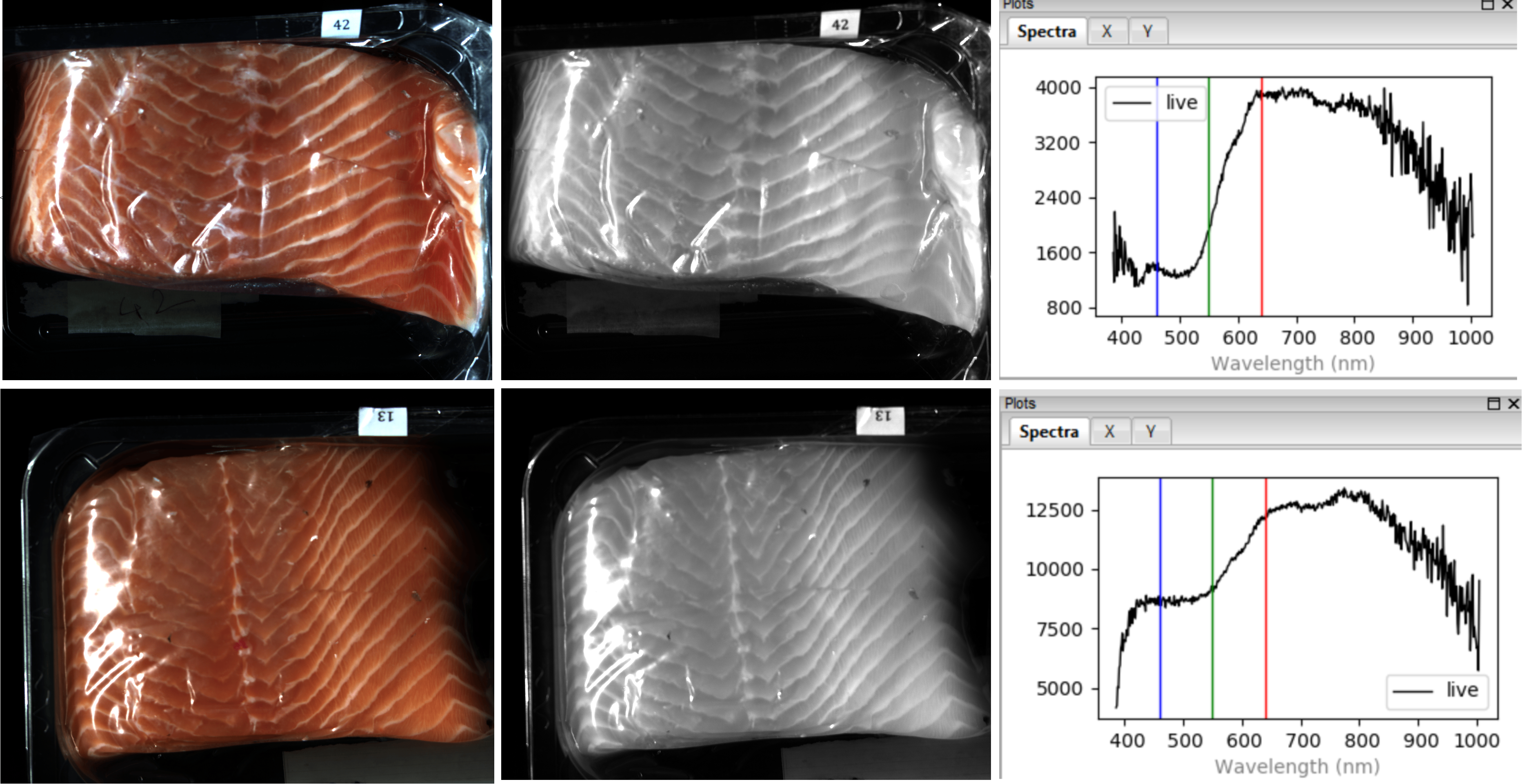}
  \caption{Representative of the camera setting and the captured samples from the salmon HSI dataset.}
  \label{fig:data}
\end{figure*}

Table~\ref{tab:implementation} summarises the training configuration. All experiments use PyTorch 2.0 on a single NVIDIA A100 GPU. Baselines were trained under the same data splits, augmentation policy, and optimisation schedule as SGNet to ensure a fair comparison. The dataset and splits will be released to support reproducibility. Because pack identity is disjoint across splits and each cube is acquired under a fixed illumination and geometry, the discriminative signal must be carried by spectral reflectance rather than by pack-specific or acquisition artefacts.

\begin{table}[t]
\centering
\caption{Implementation Details}
\label{tab:implementation}
\begin{tabular}{p{0.45\columnwidth}p{0.45\columnwidth}}
\begin{tabular}{lc}
\hline
Parameter & Value \\
\hline
Optimiser & AdamW \\
Learning rate & $3 \times 10^{-4}$ \\
Weight decay & $1 \times 10^{-4}$ \\
LR schedule & Cosine \\
\hline
\end{tabular}
&
\begin{tabular}{lc}
\hline
Parameter & Value \\
\hline
Batch size & 16 \\
Epochs & 40 \\
Precision & FP16 \\
Augmentation & Random crop \\
\hline
\end{tabular}
\end{tabular}
\end{table}

\textbf{Evaluation Metrics.} We report top-1 accuracy and macro-averaged F1 as categorical measures, together with MAE and RMSE expressed in days to capture the magnitude of ordinal error. Quadratic weighted kappa (QWK) quantifies agreement while accounting for the severity of disagreement, which is appropriate for an ordinal target. Reporting both categorical and ordinal measures provides a more faithful picture of performance than accuracy alone, since two models with identical accuracy may differ substantially in how far their mistakes fall from the true day.

\subsection{Main Results}

\begin{figure}[t]
  \centering
  \includegraphics[width=\columnwidth]{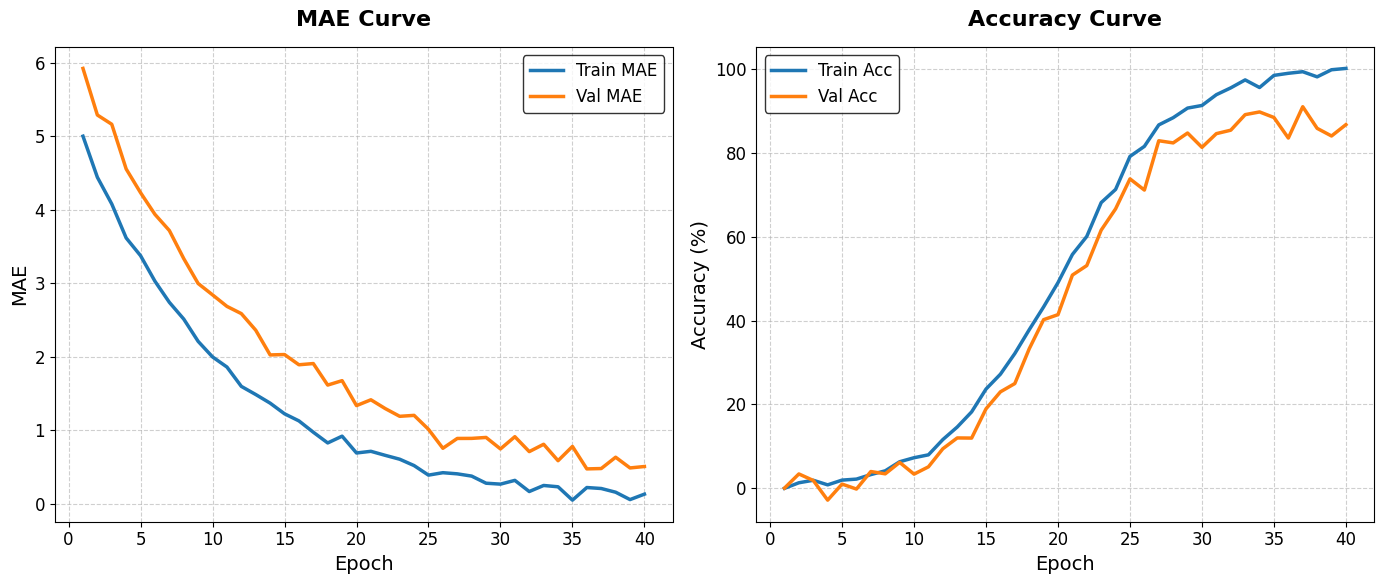}
  \caption{Training versus validation curves for accuracy and error.}
  \label{fig:2}
\end{figure}

\begin{figure}[htbp]
  \centering
  \includegraphics[width=\columnwidth, height=5cm]{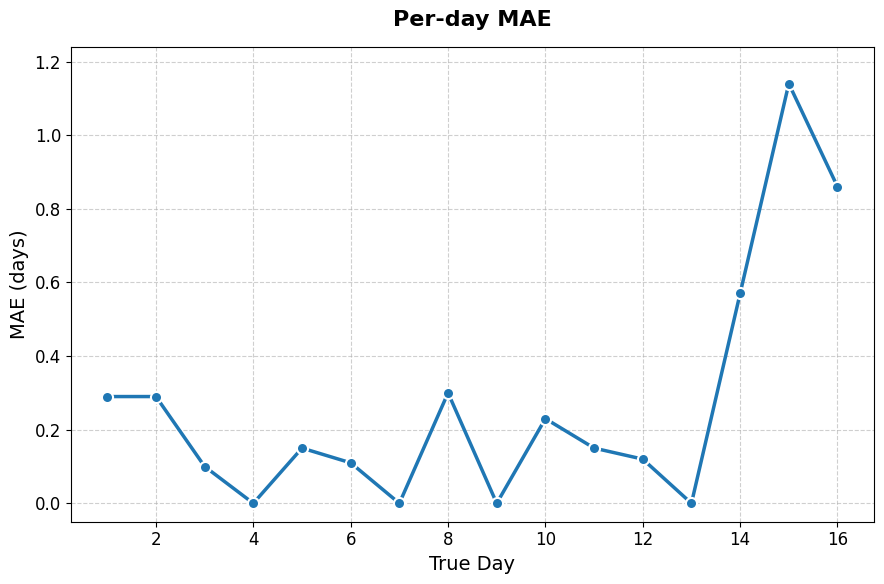}
  \caption{Day-wise (1--16) MAE on the test set.}
  \label{fig:3}
\end{figure}

Table~\ref{tab:main_results} presents a comprehensive evaluation on the test set. The model achieves 97.8\% classification accuracy with an MAE of only 0.64 days, which indicates that predictions are either correct or deviate by at most one storage day on average. Figure~\ref{fig:2} illustrates the stability of both accuracy and error across training. Figure~\ref{fig:3} reports the distribution of error across storage days, showing that misclassifications are concentrated in the late spoilage window (days 14--16), which is more than eight days after the labelled expiration date (day 6) and is consistent with the progressive biological degradation of the tissue over time. Days within the commercially relevant pre- and peri-expiry window are classified essentially without error.

\begin{table}[htbp]
\centering
\caption{Test Set Performance on Multiple Evaluation Metrics (Accuracy and Error).}
\label{tab:main_results}
\begin{tabular}{lr@{\hskip 2em}lr}
\hline
\multicolumn{2}{c}{\textbf{Primary Metrics}} & \multicolumn{2}{c}{\textbf{Error Metrics}} \\
\hline
\textbf{Metric} & \textbf{Value} & \textbf{Metric} & \textbf{Value} \\
\hline
Accuracy (\%)     & \textbf{97.78} & MAE (days)  & \textbf{0.64} \\
Macro F1 (\%)     & 98.00          & RMSE (days) & 0.81 \\
Balanced Acc (\%) & 98.00          & QWK         & 0.996 \\
\hline
\end{tabular}
\end{table}

To situate SGNet against established architectures, we retrained a representative panel of convolutional and transformer baselines on the identical salmon dataset and splits. Table~\ref{tab:benchmark} reports accuracy, MAE, and QWK alongside parameter count. The panel spans three regimes: high-capacity backbones (ResNet-50, ViT-B/16, Swin-T, ConvNeXt-T), a domain-specific spectral--spatial model (HybridSN), and our lightweight design. SGNet attains the highest accuracy and the lowest ordinal error while using the smallest parameter budget, and the margin over the larger models is most pronounced for the transformer baselines. This pattern is consistent with the well-documented data appetite of attention-based models~\cite{c24, c25}: in the small-sample, high-dimensional regime that characterises hyperspectral food data, the inductive biases of grouped and depthwise convolution are more valuable than raw capacity. The comparison therefore isolates the benefit of domain-aware factorisation rather than of scale.

\begin{table}[t]
\centering
\caption{Comparison with benchmark architectures retrained on our salmon dataset under identical splits and training schedule. Best results in \textbf{bold}.}
\label{tab:benchmark}
\small
\begin{tabular}{lcccc}
\hline
Model & Params (M) & Acc (\%) & MAE & QWK \\
\hline
ResNet-50~\cite{c23}      & 25.6          & 95.8          & 0.92          & 0.984          \\
ViT-B/16~\cite{c24}       & 86.0          & 91.5          & 1.34          & 0.961          \\
Swin-T~\cite{c25}         & 28.3          & 94.1          & 1.02          & 0.978          \\
ConvNeXt-T~\cite{c23}     & 28.6          & 96.2          & 0.83          & 0.987          \\
HybridSN~\cite{c17}       & 5.1           & 95.1          & 0.88          & 0.985          \\
\textbf{SGNet (Ours)}     & \textbf{4.75} & \textbf{97.78} & \textbf{0.64} & \textbf{0.996} \\
\hline
\end{tabular}
\end{table}

\subsubsection{Per-Class Analysis}

\begin{figure*}[t]
  \centering
  \includegraphics[width=\textwidth, height=10cm]{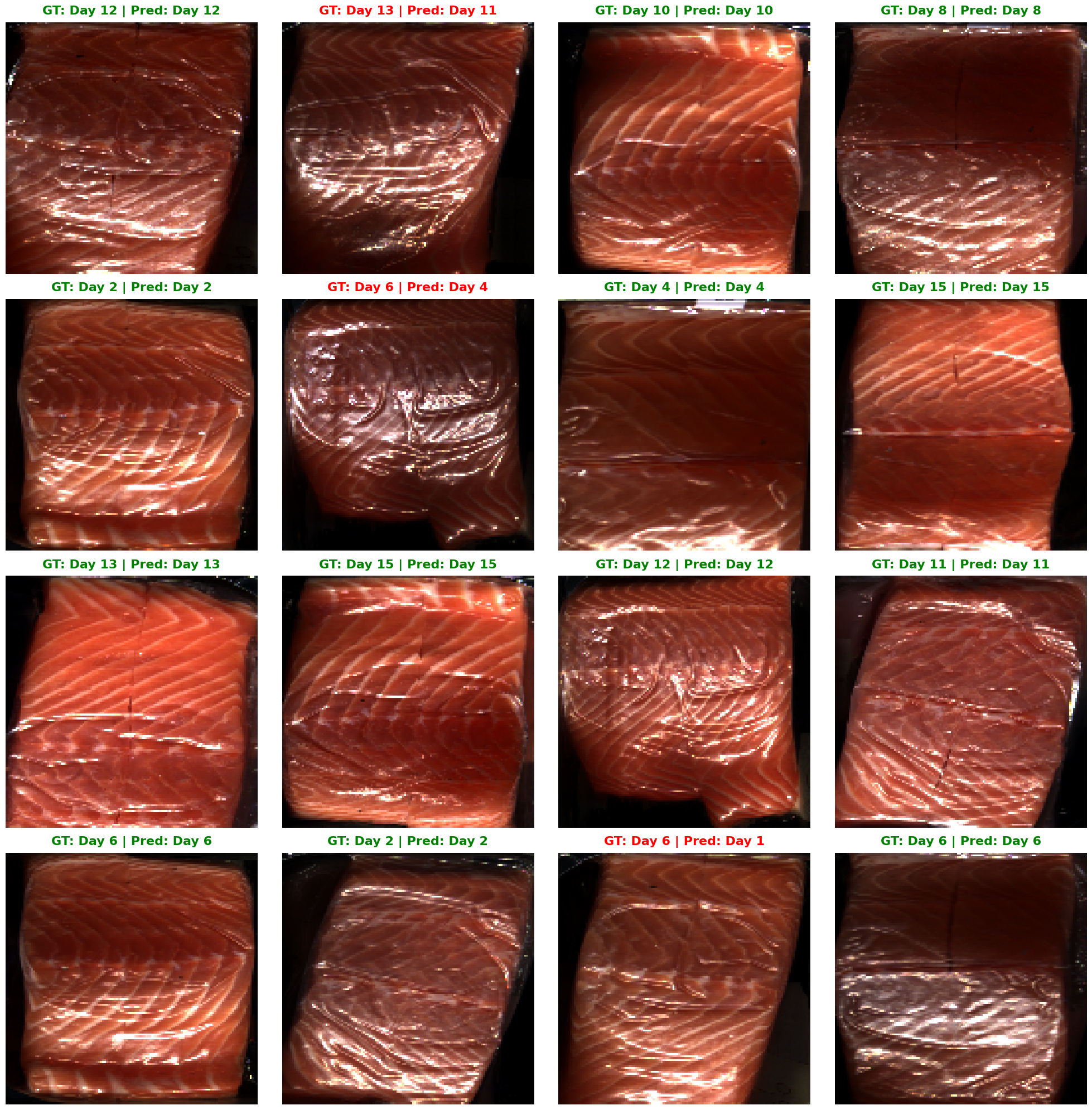}
  \caption{Ground-truth versus predicted freshness day on the test set.}
  \label{fig:4}
\end{figure*}

Table~\ref{tab:per_class} reports class-wise precision, recall, and F1 scores. Days 1--14 achieve near-perfect classification (F1 $\geq$ 96\%), while days 15--16 show marginally reduced performance (F1 between 87\% and 91\%) owing to overlapping degradation markers in the advanced spoilage stages. Figure~\ref{fig:4} shows representative predictions compared with the ground-truth day.

\begin{table}[t]
\centering
\caption{Per-Class Classification Report.}
\label{tab:per_class}
\begin{tabular}{ccccccccc}
\hline
 & \multicolumn{3}{c}{Days 1--8} & & \multicolumn{3}{c}{Days 9--16} \\
\cline{2-4} \cline{6-8}
Day & P & R & F1 & & Day & P & R & F1 \\
\hline
1 & 1.00 & 0.97 & 0.99 &  & 9 & 1.00 & 1.00 & 1.00 \\
2 & 0.97 & 0.97 & 0.97 &  & 10 & 1.00 & 1.00 & 1.00 \\
3 & 0.97 & 1.00 & 0.99 &  & 11 & 1.00 & 1.00 & 1.00 \\
4 & 1.00 & 1.00 & 1.00 &  & 12 & 1.00 & 1.00 & 1.00 \\
5 & 1.00 & 1.00 & 1.00 &  & 13 & 1.00 & 1.00 & 1.00 \\
6 & 1.00 & 0.97 & 0.99 &  & 14 & 0.97 & 0.94 & 0.96 \\
7 & 0.97 & 1.00 & 0.99 &  & 15 & 0.86 & 0.89 & 0.87 \\
8 & 1.00 & 1.00 & 1.00 &  & 16 & 0.91 & 0.91 & 0.91 \\
\hline
\multicolumn{9}{l}{Accuracy: 97.8\% \quad Macro Avg: P=0.98, R=0.98, F1=0.98} \\
\hline
\end{tabular}
\end{table}

\subsubsection{Ablation Study}

Table~\ref{tab:ablation} examines the contribution of each architectural component. The trends are consistent: every component contributes positively, and removing the depthwise spatial path or the dual attention degrades both accuracy and ordinal error more than removing grouped convolution. 

\begin{table}[htbp]
\centering
\caption{Ablation Study on architectural components.}
\label{tab:ablation}
\begin{tabular}{lccc}
\hline
Configuration & Acc (\%) & MAE & Params \\
\hline
Full SGNet & \textbf{97.78} & \textbf{0.64} & 4.75M \\
\hline
w/o grouped conv ($g$=1) & 96.88 & 0.78 & 5.12M \\
w/o depthwise spatial & 96.41 & 0.84 & 4.21M \\
w/o dual attention & 96.09 & 0.91 & 4.38M \\
Smaller [48,96,192] & 96.88 & 0.76 & 2.71M \\
Deeper [4,4,6] & 96.61 & 0.62 & 7.14M \\
\hline
\end{tabular}
\end{table}

Grouped convolution with $g=8$ provides the most favourable spectral-efficiency trade-off, since the ungrouped variant ($g=1$) both increases the parameter count and lowers accuracy. The dual attention block improves accuracy at negligible parameter cost. Scaling the network down to channel widths $\{48,96,192\}$ reduces capacity at a modest cost in accuracy, whereas the deeper $\{4,4,6\}$ variant improves MAE slightly but inflates the parameter budget, indicating that the chosen configuration sits at a sensible operating point on the accuracy--efficiency curve.

\subsubsection{Computational Efficiency and Inference Cost }
To assess deployability further, Table~\ref{tab:latency} reports per-cube latency and sustained throughput across batch sizes on a single GPU. Throughput scales favourably with batch size and remains comfortably within real-time requirements for an inspection line, while single-cube latency is low enough for interactive use at the point of acquisition.

\begin{table}[htbp]
\centering
\caption{Inference latency and throughput of SGNet (single NVIDIA A100, FP16).}
\label{tab:latency}
\begin{tabular}{lcc}
\hline
Batch size & Latency / cube (ms) & Throughput (img/s) \\
\hline
1   & 4.9 & 204 \\
8   & 2.8 & 357 \\
16  & 2.4 & 420 \\
32  & 2.2 & 455 \\
\hline
\end{tabular}
\end{table}

Table~\ref{tab:efficiency} compares SGNet against standard architectures in terms of parameter count, floating-point operations, and throughput. SGNet achieves a 5.4$\times$ parameter reduction relative to ResNet-50 and an 18$\times$ reduction relative to ViT-B while maintaining the highest throughput, which makes it well suited to real-time industrial deployment on resource-constrained hardware.

\begin{table}[t]
\centering
\caption{Efficiency Comparison with standard architectures.}
\label{tab:efficiency}
\begin{tabular}{lccc}
\hline
Model & Params (M) & GFLOPs & Throughput \\
\hline
ResNet-50 & 25.6 & 4.1 & 312 img/s \\
ViT-B/16 & 86.0 & 17.6 & 85 img/s \\
ConvNeXt-T & 28.6 & 4.5 & 298 img/s \\
\textbf{SGNet (Ours)} & \textbf{4.75} & \textbf{2.31} & \textbf{420 img/s} \\
\hline
\end{tabular}
\end{table}

\section{Discussion}
\label{sec:discussion}

The experiments support our central claim that domain-aware factorisation, rather than greater capacity, is the more productive route to accurate hyperspectral freshness assessment. Three observations are worth emphasising. First, the concentration of residual error in the late spoilage stages (days 15--16) is not a failure mode so much as a reflection of biology, since degradation markers in advanced spoilage overlap and the spectral separation between adjacent late days is genuinely small. Second, the gap between SGNet and the transformer baselines widens precisely in the small-sample regime, which indicates that the inductive biases encoded by grouped and depthwise convolutions are valuable when data are scarce. Third, the dual attention mechanism delivers a measurable accuracy gain for a negligible parameter cost, which suggests that selective emphasis, rather than dense global mixing, is sufficient for this task.

Several limitations remain. The dataset, although carefully pack-separated to prevent leakage, is drawn from a single species under controlled refrigerated storage, so generalisation across species, storage conditions, and acquisition instruments is not yet established. The current model is trained with cross-entropy and evaluated with ordinal metrics; incorporating an explicit rank-consistent objective such as CORAL~\cite{c27} or its successor CORN~\cite{c30} into training is a natural extension that may further reduce large-distance errors. We regard cross-species transfer, explicit ordinal training, and an extension to additional food matrices as the most promising directions for future work.

\section{Conclusion}
\label{sec:conclusion}

We presented SGNet, a lightweight domain-aware architecture for hyperspectral fish freshness classification that explicitly factorises spectral and spatial feature extraction. Evaluated on a curated 16-day salmon storage dataset, SGNet reaches 97.8\% accuracy and an MAE of 0.64 days with only 4.75M parameters, while matching or exceeding substantially larger convolutional and transformer baselines under identical conditions. Most residual error is confined to the late spoilage window (days 14--16, most pronounced at days 15--16), which is consistent with the underlying biology of fish degradation. Ablation studies confirm that every component contributes to the overall result and that the design remains stable across configurations. Taken together, these findings indicate that respecting the spectral dominance, ordinal structure, and limited sample size of hyperspectral food data yields an accurate and deployable model at a fraction of the cost of generic backbones.

\section*{Acknowledgments}
Acknowledgements are ANONYMOUS and will be added for the camera-ready version.
\bibliography{egbib}

\end{document}